\documentclass{elsart}

\usepackage{epsfig}

\usepackage{amssymb}

\newcommand{\phot}{photons~cm$^{-2}$~s$^{-1}$~keV$^{-1}$~}

\journal{Astroparticle Physics}

\begin{document}

\begin{frontmatter}




\title{Development of an XSPEC-Based Spectral Analysis
System for the Coded-Aperture Hard X-ray Balloon Payload EXITE2}


\author[Harvard]{P. F. Bloser\corauthref{cor}\thanksref{now}},
\corauth[cor]{Corresponding author}
\thanks[now]{Present address: Max-Planck-Institut f\"ur extraterrestrische 
Physik, Giessenbachstrasse, D-85748 Garching, Germany,
Phone: +49 89 30000-3854; Fax: +49 89 30000-3606}
\ead{bloser@mpe.mpg.de}
\author[Harvard]{Y. Chou},
\author[Harvard]{J. E. Grindlay},
\author[Harvard]{T. Narita},
\author[MIT]{G. Monnelly}

\address[Harvard]{Harvard-Smithsonian Center for Astrophysics,60 Garden
Street, Cambridge, MA 02138, USA}

\address[MIT]{Center for Space Research,
Massachusetts Institute of Technology, 
70 Vassar Street, Cambridge, MA 02139, USA}

\begin{abstract}
We present the spectral analysis system for the second-generation
Energetic X-ray Imaging Telescope Experiment (EXITE2) balloon payload.
EXITE2 is an imaging hard X-ray telescope using a coded-aperture mask
and a NaI/CsI phoswich detector operating in the energy range 20--600
keV.  The instrument was flown on a high-altitude scientific balloon
from Ft. Sumner, NM on 1997 May 7-8.  We describe the details of the 
EXITE2 spectral
analysis system, with emphasis on those aspects peculiar to
coded-aperture instruments.  In particular, we have made our analysis
compatible with  the standard X-ray spectral fitting package XSPEC by
generating a response matrix in the appropriate format including all
the effects of a coded-aperture system.  The use of XSPEC, which may
be a first for coded-aperture data, permits great flexibility in the
fitting of spectral models.  The additional effects of our phoswich
system, or any other  detector-specific considerations, may be easily
included as well.  We test our spectral analysis using observations of
the Crab Nebula, and find that the EXITE2 Crab spectrum is consistent
with those recorded by previous instruments operating in this energy
range.

\end{abstract}

\end{frontmatter}


\clearpage

\section{Introduction}
\label{sec-intro}

The hard X-ray band (20--600 keV) is a difficult energy
range for imaging instruments, since photon energies
are too high for focusing optics 
($\lesssim$ 60--100 keV, within a very small field of view) and too low
for Compton telescopes ($\gtrsim 500$ keV). 
Imaging instruments are needed to reduce source confusion in crowded
fields and perform wide field-of-view observations for sensitive
all-sky surveys.  
A practical method for hard X-ray imaging is the coded aperture
technique \cite{caroli}, in which the
position of the cosmic X-ray 
source can be deduced from the shadow 
cast by a specially-designed mask onto a large-area,
position-sensitive detector.  This technique has been successfully
applied in numerous balloon and space-based instruments using large
scintillator crystals for the detector \cite{caroli}, most notably the 
French SIGMA telescope \cite{paul}.  

We have developed the second-generation Energetic X-ray Imaging
Telescope Experiment (EXITE2) balloon payload, a
successor to the 
original EXITE instrument \cite{grindlay86}, as a coded-aperture
telescope that reduces background using the ``phoswich''  
technique \cite{kurfess}.  
Although the phoswich technique has been
used widely in non-imaging X-ray detectors, EXITE2 is an {\em
imaging} phoswich detector optimized for spatial and spectral
resolution to observe cosmic X-ray sources.  
(Another imaging phoswich balloon payload is the GRIP-2 telescope
built by Caltech \cite{schindler}.)
The EXITE2 payload has been described in detail previously 
\cite{manandhar93,lum,manandhar95,chou98}.  
Briefly, the main detector is a 36 cm $\times$ 36 cm
$\times$ 1 cm NaI(Tl) crystal surrounded on five sides by 2 cm of
CsI(Na).  Scintillation light from both crystals is collected by a $7
\times 7$ array of photomultiplier tubes (PMTs), allowing the position
of the X-ray interaction to be reconstructed by a maximum likelihood 
method (MLM) \cite{cook}.
The measured position resolution is $\delta_x =
10.6$ mm $\times (E/60{\rm keV})^{-1/2}$ (FWHM) over most of the
detector.  NaI(Tl) has a scintillation light decay time of 
230 ns, whereas the decay time for CsI(Na) is 680 ns; therefore,
the shaped pulse rise time of each event can be measured by a
pulse-shape discriminator (PSD) circuit and 
used to distinguish events absorbed in the NaI from those absorbed in
the CsI.  The energy of each X-ray event is given by the total light
recorded by all 49 PMTs (i.e., the sum of the 49 calibrated pulse heights) 
combined with the detector gain map, which
reflects that fact that photons interacting near the edge of the
detector produce
fewer detectable scintillation photons.  The energy resolution 
over the entire detector is $E_{res} \sim 
14.3\% \times (E/60{\rm keV})^{-1/2}$ (FWHM).  The PMT gains and offsets 
(that is, the pulse height channel to energy conversion) are
continuously calibrated by 
four $\sim 30$ nCi $^{241}\hspace{-.3em}$Am sources (the
``Amcal'' sources), which provide 60 keV photons, and 12 LEDs.  The Amcal 
sources are embedded in plastic
scintillator coupled to a dedicated PMT via optical fibers, allowing calibration
events to be tagged with $\sim 60$\% efficiency by detection of the 
coincident 5 MeV
$\alpha$-particle.  Since the LEDs and 
Amcal sources have known count rates, 40 counts s$^{-1}$ total for the
four Amcals and 133 Hz for the LEDs, they also provide a measure of the
detector deadtime and overall system throughput.
Imaging is achieved using a $13 \times 11$ element uniformly-redundant
array (URA) coded mask (50\% open), extended to cover $4 \times 4$
cycles, placed 2.5 m from the detector. 
The mask pixels, 
made of graded lead (12 mm), tin (1 mm), and copper (0.9 mm), are 16 mm
square, giving an 
angular resolution of 22$^{\prime}$.  A graded lead-brass collimator
limits the field of view of the detector to one cycle of the mask
pattern, or $4^{\circ}\hspace{-.4em}.65$ (FWHM).  Telescope aspect is recorded
continuously for in-flight corrections and post-flight image
reconstruction.  Elevation data is given by an inclinometer and
shaft-angle encoder, while azimuth data during the daytime are given
by either a sun sensor or differential GPS system.  We found these
daytime aspect
systems to be accurate to $\sim 6^{\prime}$.  At night an intensified
video camera records star fields on-axis, allowing pointing to be
determined to $\sim 1^{\prime}$.  

In this paper we report on calibration observations of the Crab Nebula
performed by EXITE2 during its first science flight from Ft. Sumner,
NM on 1997 May 7--8.  
We focus on our spectral analysis system, 
which we have made compatible with the standard X-ray spectral analysis
software package XSPEC \cite{arnaud}.  This may be the first use of XSPEC
with a coded-aperture telescope, and so we describe in detail the method by which
our data are analyzed, including the effects of the coded-aperture imaging system
in the response matrix.  Such a system could be made to work as well for future
coded-aperture instruments, such as INTEGRAL and the proposed hard X-ray imaging
survey telescope EXIST \cite{grindlay2000}.

\section{Observations}
\label{sec-obs}

The EXITE2 payload was launched from the National Scientific Balloon
Facility (NSBF) base in Ft. Sumner, NM at UT 16:15 on 1997 May 7.  Due
to high westerly winds at our float altitude (115,000--127,000 feet),
the observing time was limited to 15 hours.  During this time
observations were performed of the Crab Nebula, the X-ray binaries 4U
0614+09 and Cyg X-1, and the active galactic nuclei (AGN) 3C 273 and NGC
4151.  In this paper we describe our Crab observations; our
observations of Cyg X-1 and the AGNs are discussed in separate 
papers \cite{chou2000,chou2001}.  
We did not detect 4U 0614+09, though our data amounted to only 101
seconds of live time (see below).  

The Crab Nebula was observed first in the flight, from UT 23:16:00 on
1997 May 7 until UT 
00:16:00 on 1997 May 8 (17:16:00--18:16:00 local time).  Several
problems occurred during the Crab observation that limited the
effective observing time.  During the initial daytime portion of the
flight 
the azimuth gyroscope electronics overheated, so that the Crab (and 4U
0614+09) drifted in and out of the field of view.  
(This was fixed by power cycling later in the day and was not a
problem during the night.)
In addition, a
discriminator was incorrectly set, resulting in a large deadtime.
Nevertheless, the Crab was strongly detected in only $\sim 5$ minutes
of live time, as measured by the recorded Amcal rate.

\section{EXITE2 Spectral Analysis Procedure}
\label{sec-proc}

The analysis of EXITE2 spectra is performed with XSPEC.  The XSPEC
program requires as input two files in the FITS format: a Pulse
Height Analyzer (PHA) file, 
containing the count spectrum, and a Redistribution Matrix File (RMF),
which describes the instrument response and the pulse height to energy
conversion.  Additional information may
be provided in an Ancillary Response File (ARF), which contains
additional multiplicative factors for the response as a function of
energy.  We note that the standard XSPEC Background (BKG) file is not 
needed in our case, as the source counts derived from a coded-aperture
telescope are already background-subtracted (see Section~\ref{sec-pha}).
The standard practice is to include in the RMF only the relative
probabilities, normalized for each matrix row, of photons being recorded 
in various pulse height channels (that is, ``redistributed'') due
to the detector energy resolution, escape peaks, Compton scattering, and any 
other effects.  The effective area as a function of energy is then contained in
the ARF file, which includes the effects of the detector efficiency, absorbing 
materials, collimator response, etc.  In the case of EXITE2, however, the 
inclusion of the PSD rejection efficiency makes the calculation
of the redistribution and the effective area difficult to disentangle (see 
Section~\ref{sec-rmf}).  We have therefore strayed from convention and 
placed the effective area directly in the RMF file, including all effects
which do not change from observation to observation.  We have used the ARF file
to contain observation-dependent corrections to the effective area, 
specifically the effects of atmospheric absorption and collimator transmission
due to pointing errors.  

\subsection{Count Spectrum: The PHA File}
\label{sec-pha}

The PHA file contains the count spectrum, or the number of counts
detected in a given time within each detector pulse height channel.
Once the detector is calibrated, these pulse height channels correspond
directly to energy bands.
Since EXITE2 is a coded-aperture imaging telescope, the
counts in each channel are found by producing an image in each of the 12
logarithmically-spaced EXITE2 energy bands (chosen to contain roughly 
equal counts from a Crab-like
spectrum).  Images are formed by correlating the recorded detector image
(the shadow of the mask) with the mask pattern; for a uniformly-redundant mask,
the result for a point source is a correlation
peak, containing the source counts, on top of a flat background.
(Details of coded aperture image reconstruction are
given by Caroli et al.  \cite{caroli} and references therein; see also 
\cite{covault} and  \cite{manandhar95}.)
In the EXITE2 implementation, the detected counts within each band were 
binned spatially into
$162 \times 162$ ``subpixels'' to form detector images.  The subpixel
size, 2.7 mm or 3.7$^{\prime}$, was chosen to be one-sixth of the mask
pixel size, allowing the 
mask pattern to be well over-sampled.  Because the detector area is not equal 
to an integral number of mask patterns ($\sim 3.6$), it was then necessary
to ``pixel shuffle'' the detector images in order to perform the correlation
using the whole area.
This means that counts from the entire
detector were shifted and averaged into 
the corresponding subpixels of one basic mask pattern, in the process improving
the signal-to-noise\cite{covault,chou2000}.  
These shuffled detector images were
correlated with the mask pattern to produce sky images.  The source
counts were then given by the maximum value within an $11 \times 11$
subpixel (40.7$^{\prime}$) box around the correlation peak.  The mean 
background level and RMS noise were
calculated using all counts outside this box, and the mean was
subtracted from the peak value.  
The EXITE2 correlation is 
``balanced,'' meaning that the numerical mask pattern is composed of
1's and -1's to represent, 
respectively, open and closed mask pixels; this mean background level should
therefore be close to zero \cite{fenimore}.

Due to the poor pointing stability during the
Crab observation, it was necessary to produce images in each band in
one second increments.  These images were then shifted and added
according to the aspect solution from the shaft angle encoder
(elevation) and differential GPS (azimuth, which agreed with the
independent azimuth determination provided by the sun sensor), since 
the Crab was
observed during the day.  Only images with an offset less than 20
subpixels ($\sim 1^{\circ}\hspace{-.4em}.2$) were included.  For each
offset included the collimator response was calculated, and a total
weighted average of the collimator attenuation was included as a
correction in the ARF file (see Section~\ref{sec-arf}).  
The stacked images thus
produced give the total source counts in each pulse height band; these 12
numbers and their errors make up the PHA file.  The total live time, as
calculated from the total number of Amcal events in all the data
used in the stacked images, is also included in the PHA file.  
Finally, an ``area scaling factor'' is needed in the PHA file due to the
pixel shuffling procedure: since the counts in corresponding subpixels are
averaged, not added, the total counts appearing in the correlation
peak will be reduced from the total recorded (although
the averaging improves the signal-to-noise\cite{covault,chou2000}).  
This scaling factor is simply the ratio of the area
of the basic mask pattern to that of the entire detector.

\subsection{Response Matrix: The RMF File}
\label{sec-rmf}

The response matrix describes how photons of a given energy
are recorded as counts within a given detector pulse height channel, and
gives the conversion between pulse height channel and energy.  The physics 
of photon interactions in the 
detector material and all surrounding materials must be included.
In addition, any further effects peculiar to a given instrument must
be taken into account.  In a coded-aperture telescope, an important 
correction arises due to the finite position resolution of the detector.  
Since the recorded detector image will be blurred by the position
resolution, the height of the correlation peak will be 
reduced \cite{caroli,covault,manandhar95}.  In EXITE2, this
effect is energy and position dependent, since the position resolution 
depends on the number of scintillation photons collected. 
In addition, residual glow in the detector, caused mainly by long time-constant
decay modes in the CsI after cosmic ray interactions, leads to an
additional degradation in position resolution \cite{chou98}.  
Thus it is necessary
to calculate the appropriate ``imaging factors'' for each energy band.  This 
was done by convolving the ideal detector image (the mask pattern) with a
Gaussian whose width was given by the measured spatial resolution and varied
spatially according to the gain map.  The effect of residual glow, as
measured by the position resolution of the calibration LED 
images \cite{chou98},
was included as well.  This blurred detector image was then
correlated with the mask pattern.  The ratio of the height of the
correlation
peak to the height of the mask auto-correlation peak 
gives the imaging factor.  The EXITE2 imaging factors are plotted as
a function of energy in Figure~\ref{fig-caroli}.
An additional minor correction in 
coded-aperture telescopes may be made due to the transparency of the 
mask at high energies, which reduces the difference between open and
closed pixels in the detector image.  This is not a large effect in the
EXITE2 energy range, but we include it.

Additional corrections depend on the type of detector involved.  In the
case of a phoswich detector such as EXITE2 the PSD rejection efficiency, 
or the probability that an event which deposits energy in the CsI will 
be rejected, plays a key role and must be included.
This effect, which may be energy-dependent, 
helps determine the effective area of the detector, since accepting all
CsI events would mean including two additional centimeters of sensitive
volume below the NaI.
Other effects could include the probability of 
background-rejection techniques excluding valid events.  For example,
EXIST will include depth-sensing at low energies to reduce background, and
so the effective detector thickness will be affected \cite{grindlay2000}.

We constructed the EXITE2 response matrix
using the Monte Carlo simulation package MGEANT \cite{sturner} combined with
extensive laboratory calibrations using radioactive sources.  
MGEANT is based on the CERN Program Library particle
propagation and interaction simulation
package GEANT, providing an improved user interface for specifying
detector materials and geometry.  We produced a detailed simulation of
the EXITE2 detector and all surrounding materials, including the mask,
collimator, entrance windows, insulation, NaI, CsI, and aluminum
housing.  The MGEANT EXITE2 model is shown in Figure~\ref{fig-schem}.
Then, at a series of input energies between 10 keV and 1200 keV, 
$2.5 \times 10^4$ monoenergetic photons 
were allowed to enter the system.   
The input energies were spaced roughly logarithmically between 10 and
600 keV, giving 7--8 inputs per EXITE2 band, and every 30 keV from 600
keV to 1200 keV.
For each input photon the total
energy deposited by all interactions
in the NaI and CsI crystals was recorded.  The
recorded energies were first blurred by a Gaussian distribution corresponding
to the measured detector energy resolution (Section~\ref{sec-intro}; this capability
is included in MGEANT) and then
binned into the 12 EXITE2 energy bands.
The pulse height boundaries of the energy bands were determined
based on 49-point scans with radioactive line sources, in which the 
response of each PMT was recorded.
If more than half of the total energy was deposited in the CsI then the event
was accepted or rejected according to the measured PSD efficiency for that
band.  This selection criterion is somewhat arbitrary, and could be improved
with more careful modeling of the pulse shape expected from a combined NaI/CsI 
event.  No correction was made for the slightly lower ($\sim 85$\%) effective light
yield of the CsI.  

Since the deposition of energy in the CsI is related to the 
probabilistic redistribution of photon energies (through Compton scattering,
emission of K-escape photons, etc.) but the vetoing of events based on
the PSD rejection efficiency helps determine the effective area, it was not
possible to separate the two effects into separate files, as pointed out
above.  Therefore all unchanging factors which determine the effective area
were placed in the main response matrix, or RMF file.  The coded mask and
other absorbing materials were already present in the MGEANT model.
The fraction of the $2.5 \times 10^4$ input events recorded in
each band 
times the detector geometric area
gives the effective area in that band.  These effective areas were
further reduced by the appropriate imaging factors and mask
transparency factors.  The results are shown in Figure~\ref{fig-eff}.  
The resulting response matrix 
was finally converted into a FITS RMF file for use in XSPEC.

\subsection{Atmospheric Attenuation and Collimator Transmission: The ARF File}
\label{sec-arf}

As a balloon-borne instrument, EXITE2 suffers from atmospheric 
attenuation and pointing errors.
Since the attenuation and pointing history change from observation to
observation, we have not included them in the RMF file.  Rather, we have
created an ARF file for each observation containing only the
additional multiplicative factors appropriate for the atmospheric
grammage at 
that source elevation and balloon altitude and the weighted collimator
response of the one-second intervals included (see Section~\ref{sec-pha}).  
(The altitude varied between
115,000 
feet and 127,000 feet, or $\sim 6$ g cm$^{-2}$ and $\sim 3.5$ g
cm$^{-2}$ of residual atmosphere, during the flight.)
As the source elevation
changes during the course of an observation, the entire grammage
history was taken into account and a weighted average attenuation at
each energy was calculated.  These multiplicative factors were combined 
at the energy of each response matrix row and converted into a FITS ARF file.

\section{Results}
\label{sec-res}

The effective area of the EXITE2 instrument, as calculated using
MGEANT and recorded in the response matrix including all the effects
described in Section~\ref{sec-rmf}, is shown as a function of
energy in Figure~\ref{fig-eff}.  This figure includes the effects of
the residual atmosphere and pointing errors (recorded in the ARF file) 
from the Crab observation.  
The response remains flat above 100 keV due to 
decreased PSD rejection efficiency above $\sim 250$ keV, which
effectively increases the 
thickness of the detector (and increases the background, of course).  
A detailed discussion of the PSD efficiencies and the actual
sensitivity versus energy are given in a separate paper \cite{chou2000}.
(We note that a previously-published calculation of the EXITE2 effective
area \cite{lum} used overly-optimistic values of the imaging factors and
did not include the effects of PSD rejection efficiency.)
Figure~\ref{fig-eff} provides a vivid illustration of the difficulties
inherent in performing sensitive imaging observations in the hard X-ray band.
Although the geometric area of the NaI crystal is 1296 cm$^2$, at
100 keV the degrading factors of the mask (0.5), atmosphere (0.49),
collimator on-axis transmission (0.84), pointing errors (0.83), 
imaging factor (0.55), and other absorbing material conspire to reduce 
the effective area to about 115 cm$^2$.

As described in Section~\ref{sec-obs}, the live time during the Crab
observation was limited by poor aspect and an improperly set
discriminator to just 297 seconds.  Nevertheless, the Crab is clearly
detected.  In Figure~\ref{fig-image} we show the EXITE2 image of the
Crab Nebula in the 50--70 keV band, in which the source is detected at a
significance of $5.7 \sigma$.  The contours begin at $3 \sigma$ and
represent steps of $0.5 \sigma$.  The peak contour is located $\sim
10^{\prime}$ from the true location of the Crab, indicated by the
cross.  This is slightly greater than the $6^{\prime}$ accuracy of the
differential GPS azimuth determination and represents the uncertainty in the
boresighting between the GPS and intensified star camera.

The Crab was detected in EXITE2 images between 30 and 300 keV.  We
formed a count spectrum as described in Section~\ref{sec-pha} and fit
it with a single power law model using XSPEC.  In
Figure~\ref{fig-spec} we show the EXITE2 Crab spectrum with the best fit
power law of the form $F = C(E/100 {\rm keV})^{-\alpha}$ \phot (the
{\em pegpwrlw} model in XSPEC, which allows the user to select the
normalization energy).  At
the top is the raw count spectrum, in the middle are 
the fit residuals, and at the bottom is the unfolded source photon
spectrum.  
The $\sim 2\sigma$ residual in the lowest energy bin could be due to
an incorrect value for the PSD rejection efficiency; at low energies
the effective rise times for NaI and CsI events are similar, making good
discrimination difficult \cite{chou98,chou2000}.
The best fit parameters are photon index $\alpha = 2.36 \pm
0.33$ and normalization $C = (6.62 \pm 1.12) \times 10^{-4}$ \phot at
100 keV (errors are 90\% confidence for one interesting parameter).
The reduced $\chi^2$ was 0.86 for 6 degrees of freedom.  A 
broken power law model did not improve the fit, and XSPEC could not
constrain the break energy.  

In Figure~\ref{fig-specs} we compare the EXITE2 Crab spectrum with
results from other instruments operating in the hard X-ray band,
including GRIS \cite{bartlett}, HEAO A-4 \cite{jung}, and OSSE \cite{much}.  
Both the GRIS and HEAO A-4 Crab spectra are best fit with
broken power laws (a best fit for OSSE is not reported), 
though the break energies are quite different: $E_b = 60$ keV for
GRIS and $E_b = 127.7$ keV for HEAO A-4.  The power law indices reported above
their respective break energies for 
GRIS ($\alpha_2 = 2.22$) and HEAO A-4 ($\alpha_2 = 2.48$) are consistent
with the EXITE2 power law index, though at lower energy the spectra
flatten ($\alpha_1 = 2.00$ for GRIS, $\alpha_1 = 2.08$ for HEAO A-4).
Even these values are consistent with the EXITE2 spectral index within
$2\sigma$.  The GRIS, HEAO A-4, and OSSE normalizations near 100 keV
only agree to within 20\%; for GRIS and HEAO A-4 this is a $\sim
7\sigma$ discrepancy \cite{bartlett}.  Our short observing live time does
not allow us to resolve this issue; the EXITE2 data are consistent,
within the error bars, with all previous measurements in the hard
X-ray band (Figure~\ref{fig-specs}).

\section{Conclusion}

We have successfully generated a response matrix for the EXITE2
imaging hard X-ray telescope and demonstrated its accuracy through
calibration observations of the Crab Nebula.  The peculiar effects of
an imaging phoswich detector using a coded aperture mask have been
accounted for and included.  Our spectral analysis system is
compatible with XSPEC, making it extremely flexible for detailed
spectral fitting.  
We shall incorporate XSPEC into further EXITE2 analysis and suggest it
be used for future coded-aperture analysis (e.g. with INTEGRAL or EXIST).
The EXITE2 results for the Crab nebula are
consistent with measurements by other instruments in the same energy
band.

We wish to thank T. Gauron, J. Gomes, V. Kuosmanen, F. Licata,
G. Nystrom, R. Scovel, and J. Apple for technical support, and
J. Grenzke, K. Lum, and B. Robbason for software development.
This work was
supported in part by NASA grants NAGW-624 and NAG5-5103.
PFB acknowledges partial support from
NASA GSRP grant NGT5-50020.

\newpage

\begin{figure}[t]
\epsfig{figure=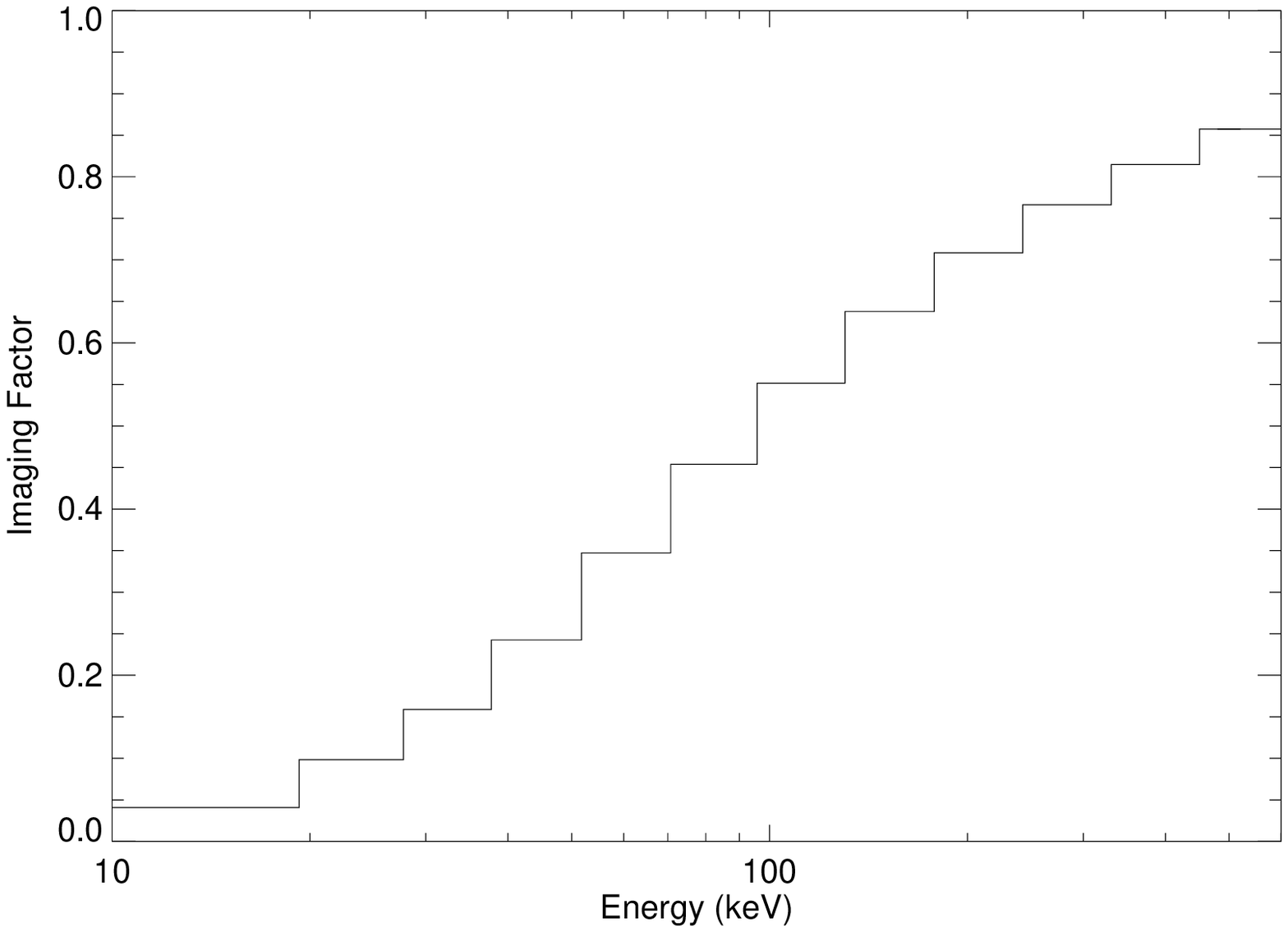,height=3.5in}
\caption{\label{fig-caroli}}
\end{figure}

Fig~\ref{fig-caroli}: The calculated imaging factors for the EXITE2 
coded-aperture telescope as a function of energy.  The factors arise due
to the finite position resolution of the detector, which degrades the
correlation peak.  All factors which affect the position resolution, 
including positional gain variations and residual glow from cosmic
ray interactions, have been included.

\begin{figure}[t]
\epsfig{figure=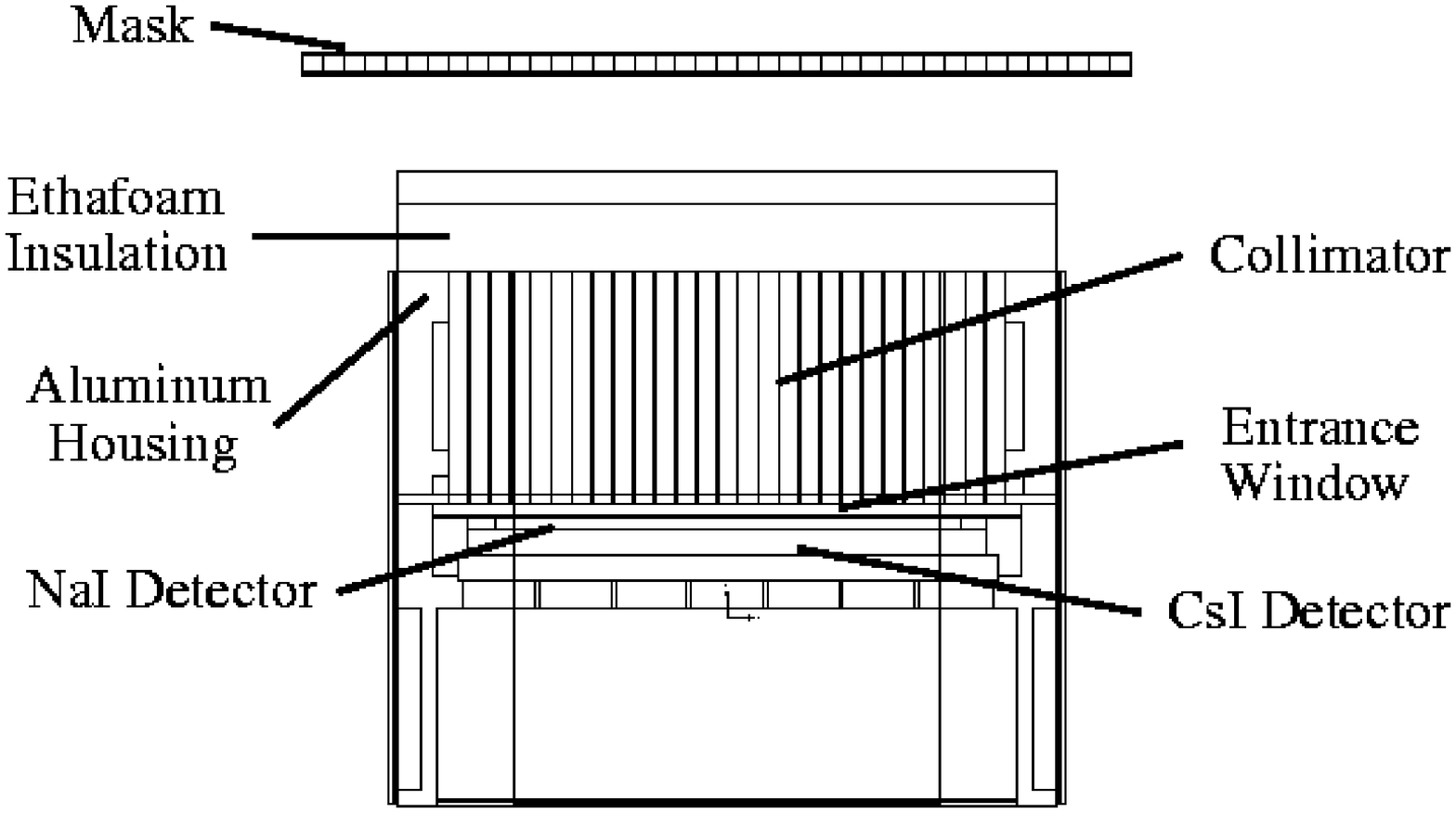,height=3.5in}
\caption{\label{fig-schem}}
\end{figure}

Fig~\ref{fig-schem}: MGEANT model of the EXITE2 detector.
The components affecting the response matrix are labeled, including
the coded mask, collimator, mylar and aluminum entrance 
windows, NaI and CsI crystals, aluminum housing, and Ethafoam thermal
insulation.

\begin{figure}[t]
\epsfig{figure=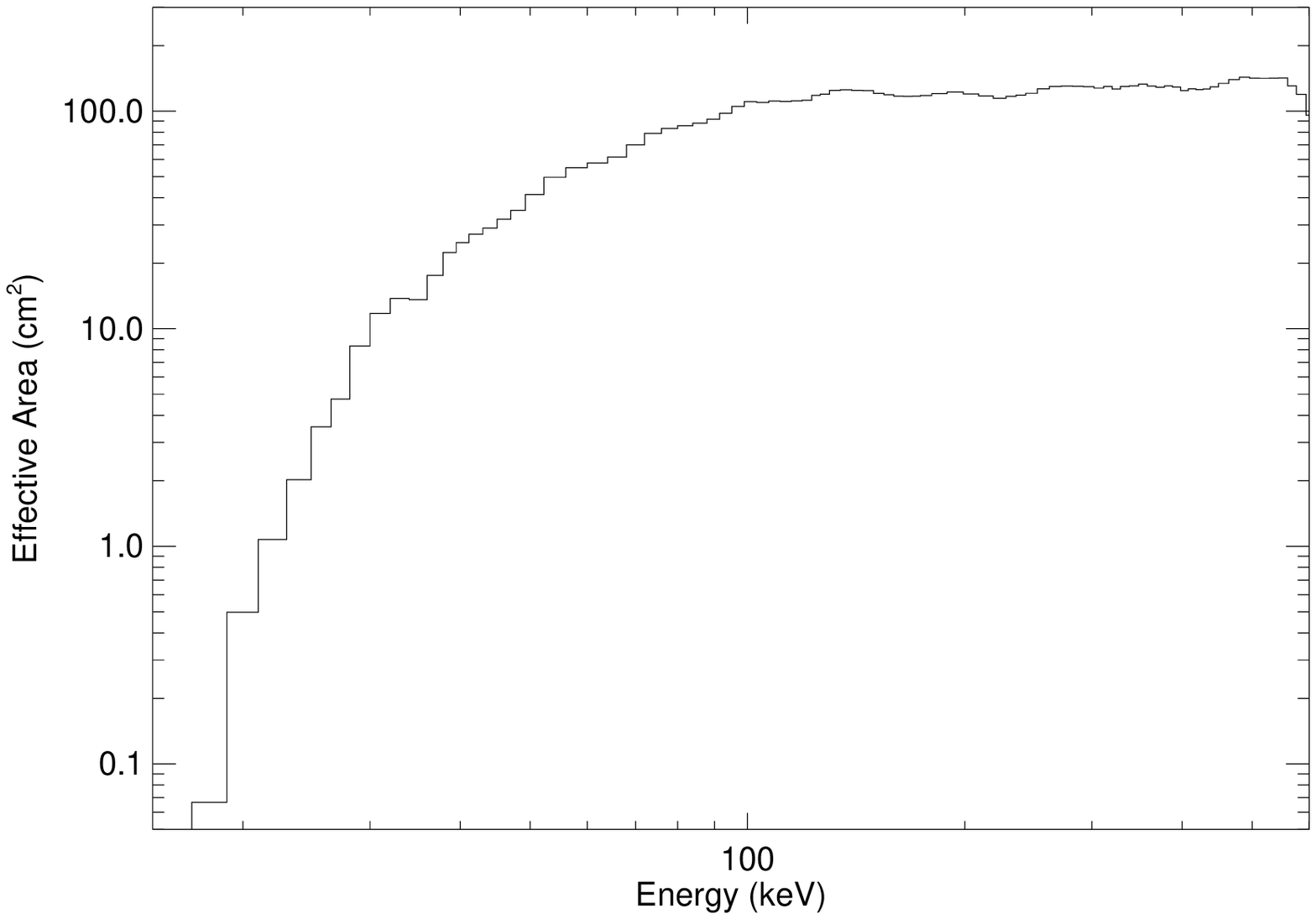,height=3.5in}
\caption{\label{fig-eff}}
\end{figure}

Fig~\ref{fig-eff}: Effective area of the EXITE2 detector from
MGEANT simulations.  The atmospheric and pointing parameters of the Crab
observation were used.  Included are the effects of mask attenuation,
PSD rejection efficiency, mask transparency, and imaging factors.
The response remains flat above 100 keV due to
decreased PSD rejection efficiency, which effectively makes the
detector thicker.

\begin{figure}[t]
\epsfig{figure=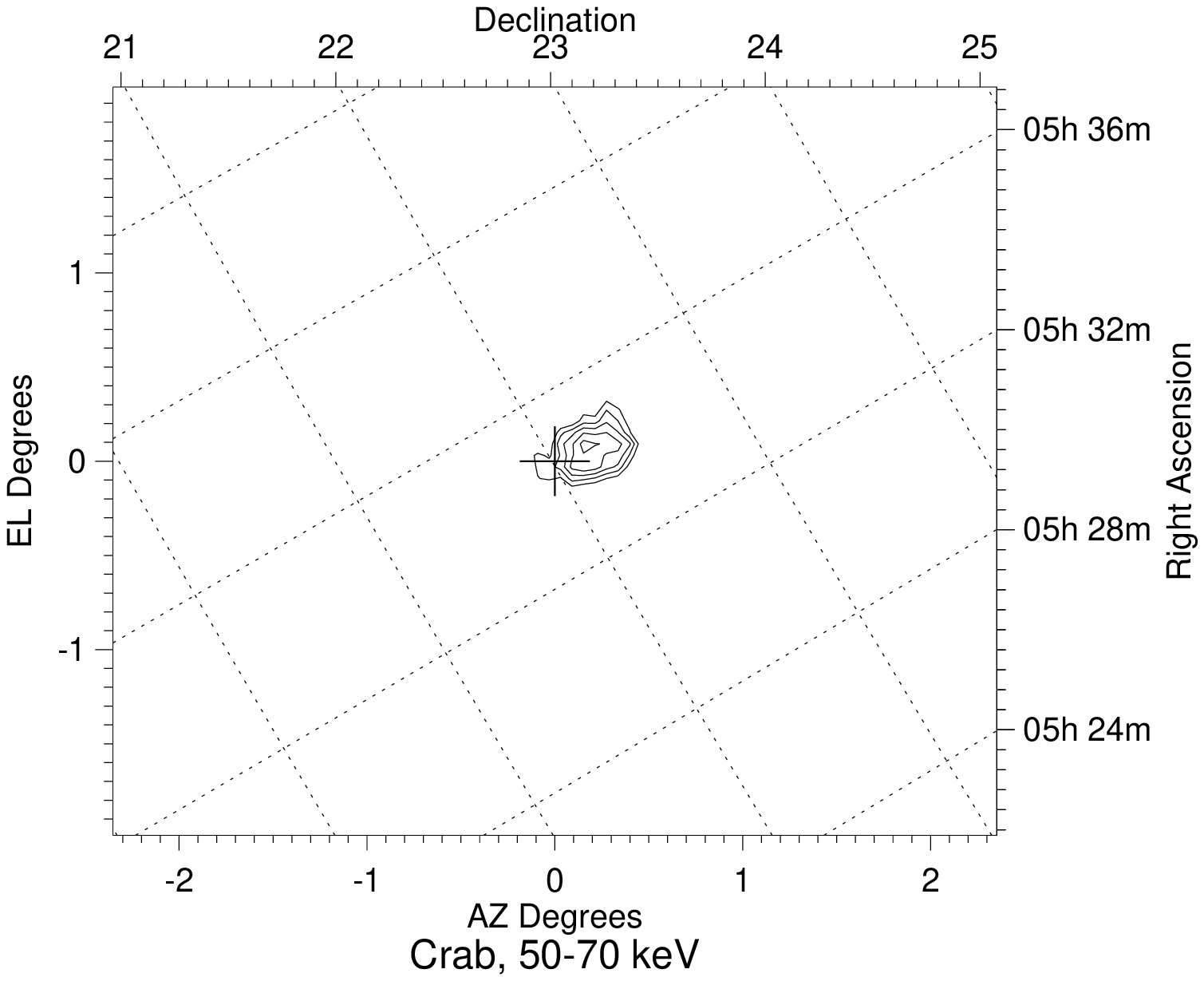,height=3.5in}
\caption{\label{fig-image}}
\end{figure}

Fig~\ref{fig-image}: EXITE2 image of the Crab Nebula, 50--70
keV.  The source is detected at a significance of 5.7 $\sigma$ in this
range despite a live time of only 297 seconds.  The contours begin at
3 $\sigma$ and represent steps of 0.5 $\sigma$.

\begin{figure}[t]
\epsfig{figure=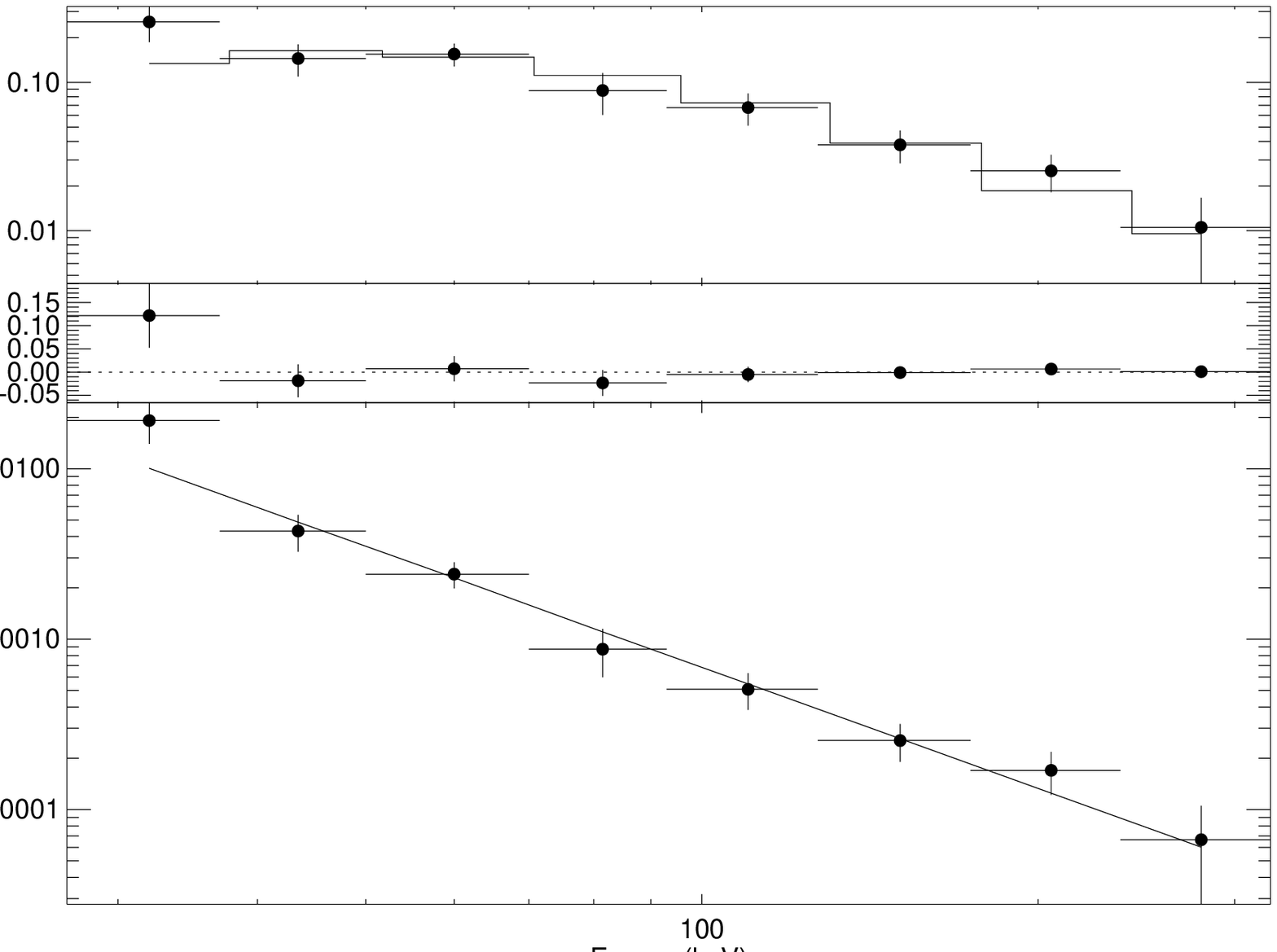,height=3.5in}
\caption{\label{fig-spec}}
\end{figure}

Fig~\ref{fig-spec}: EXITE2 spectrum of the Crab with a
power law fit.  The raw count spectrum is shown at the top, the fit
residuals in the middle, and the unfolded photon spectrum at the
bottom.  The fit parameters are given in the text.

\begin{figure}[t]
\epsfig{figure=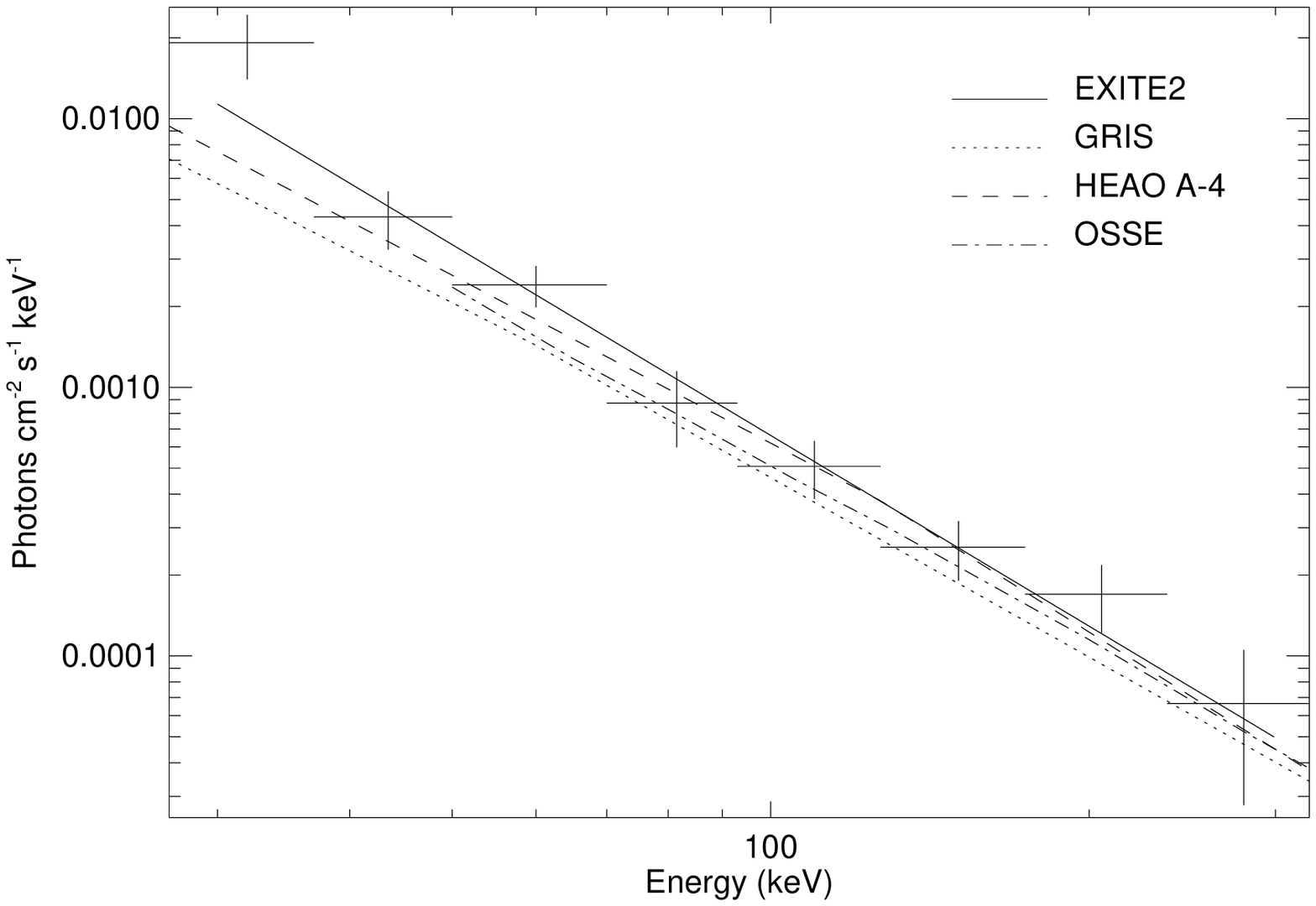,height=3.5in}
\caption{\label{fig-specs}}
\end{figure}

Fig~\ref{fig-specs}: The best-fit EXITE2 power law 
spectrum of the Crab,
together with Crab spectra from GRIS, HEAO A-4, and OSSE.  Also shown
are the unfolded EXITE2 data points.

\clearpage

\clearpage

\clearpage

\clearpage

\clearpage

\clearpage

\end{document}